\documentclass[reprint,singlecolumn]{revtex4}
\usepackage{amsfonts}
\usepackage[tbtags]{amsmath}
\usepackage{amssymb}
\usepackage{amsthm}
\usepackage{color}
\usepackage{charter}
\usepackage{graphicx}
\usepackage{lipsum}

\begin{document}

\title{Photon blockade in a bi-mode nonlinear nano-cavity embedded with a
quantum-dot}
\author{Xinyun Liang}
\author{Zhenglu Duan}
\email{duanzhenglu@jxnu.edu.cn}
\author{Qin Guo, Shengguo Guan, Min Xie, Cunjin Liu}
\address{{College of Physics and Communication Electronics, Jiangxi Normal
University, Nanchang 330022, China}}

\begin{abstract}
We study the interaction between a quantum-dot and a bi-mode micro/nano-
optical cavity composed of second-order nonlinear materials. Compared with
the Jaynes-Cummings (J-C) model, except for a coherent weak driving field, a
strong pump light illuminates the two-mode optical cavity. Analytical
results indicate that the model exhibits abundant non-classical optical
phenomena, such as conventional photon blockade induced by the nonlinear
interaction between polaritons. It constitutes unconventional photon
blockade induced by quantum interference due to parametric driving. We
compare the photon statistical properties and average photon number of the
proposed model, J-C model, and double-mode driven optical cavity under the
same parameters and the proposed model can obtain stronger antibunching
photons and higher average photon number.
\end{abstract}

\maketitle

\section{Introduction}

Nonclassical light generation is an important objective in the fields of
quantum optics, quantum information, and quantum computation\cite{1,2,3,4,5}%
. Intensity fluctuation of the light field can be used to identify the
quantumness of light, e.g., $g^{\left( 2\right) }\left( 0\right) \geq 1$
represents classical light, whereas $g^{\left( 2\right) }\left( 0\right) <1$
denotes nonclassical light. Photon blockade (PB) is considered an important
mechanism to generate nonclassical light. When a coherent laser imprints on
a nonlinear quantum optical cavity, a photon existing in the cavity may
blockade other photons entering the cavity because of the strong interaction
between photons\cite{6}. In the limit of strong nonlinearity, the PB effect
could be a good mechanism to generate single photons with perfect purity.
The PB effect, which is based on strong nonlinearity, has been studied in
many quantum systems, such as atom-cavity system\cite{7}, single quantum-dot
cavity system\cite{8,8a}, $\chi ^{\left( 2\right) }$ nonlinear doubly
resonant nanocavity system\cite{13a}, and opto-mechanical systems\cite%
{15,15a,15b}.

Recently, researchers have found that a strong antibunching light can be
obtained by the quantum interference of different driven-dissipative
pathways in a quantum optics system\cite{6a}. Contrary to the strong
linearity induced PB, weak nonlinearity helps realize the photon
antibunching induced by quantum destructive interference. To differentiate
the photon antibunching processes induced by different physical mechanisms,
the antibunching induced by strong nonlinear interaction between photons is
referred to as conventional photon blockade (CPB), whereas that by the
quantum interference is termed unconventional photon blockade (UCPB). Since
its proposition, many theoretical and experimental works have focused on
UCPB in different systems, such as coupled nonlinear cavities\cite{6a},
superconducting qubit microwave cavity\cite{11} and opto-mechanical systems,
two coupled nonlinear single-mode cavity system\cite{13}, and $\chi ^{\left(
2\right) }$ nonlinear coupled or doubly-resonant cavity system\cite{14,14a}.

Currently, studies have found that the quantum interference phenomenon can
be used to enhance the CPB, and consequently the purity of the generated
single-photon\cite{10}. In addition, we note that CPB and UCPB can coexist
in the same quantum optics system\cite{10,our-work}, which is helpful in
further understanding the physical mechanism behind CPB/UCPB.

In this study, we propose a model to generate strong PB with a single
quantum-dot (QD) that is strongly coupled to a bi-mode nonlinear nanocavity,
which is filled with $\chi ^{\left( 2\right) }$ medium. The proposed model
can be considered a composite model of the J-C model and the bi-mode driven
nonlinear optical cavity. Studies have indicated that CPB and UCPB exist in
the Jaynes-Cummings (J-C) model with strong coupling and the bi-mode driven
nonlinear optical cavity\cite{14,14a}. Thus, we expect CPB induced by strong
nonlinear interaction and UCPB induced by quantum interference to
simultaneously appear in the proposed model, when the system is in the
strong coupling regime. By studying the proposed model, we find that strong
UCPB appears in the red atomic detuning region, which does not occur in the
J-C model. Comparing the J-C model with bi-mode driven nonlinear nanocavity,
more transition pathways from ground to two-photon states in the proposed
model make the quantum statistics of the photons more complicated.

The remainder of this paper is organized as follows. In Section II, we
present the effective Hamiltonian to describe the proposed system. In
Section III, we study the physical mechanism of PB using the wavefunction
approach, and provide the optimal conditions for strong photon antibunching
via second-order correlation function. In addition, we theoretically study
the relationship between the PB effect in our model and J-C and empty cavity
models using the quantum interference transition paths. In Section IV, we
provide the numerical results of a second-order correlation function and
compare them with the analytical results. Finally, we conclude this paper in
Section V.

\section{Model}

The proposed model is depicted in Fig. 1(a) with a single QD embedded in a
doubly resonant nonlinear nanocavity made of $\chi ^{\left( 2\right) }$
material. For $\chi ^{\left( 2\right) }$ nonlinearity, the fundamental and
the second harmonic modes are coupled together and mediate the conversion of
one photon in second-harmonic mode to two photons in fundamental mode. In
the rotating frame of the weak driving and strong pump lights defined by $%
U\left( t\right) =\exp \left[ -i\omega _{d}\left( \sigma _{+}\sigma _{-}+%
\hat{a}^{\dagger }\hat{a}\right) t-i\omega _{p}\hat{b}^{\dagger }\hat{b}t%
\right] $, the interaction Hamiltonian of the system is given by,
\begin{eqnarray}
H &=&\delta \sigma _{+}\sigma _{-}+\Delta _{a}\hat{a}^{\dagger }\hat{a}%
+\Delta _{b}\hat{b}^{\dagger }\hat{b}+E\left( \hat{a}+\hat{a}^{\dagger
}\right) +F\left( \hat{b}+\hat{b}^{\dagger }\right)  \label{H} \\
&&+g\left( \sigma _{+}\hat{a}+\hat{a}^{\dagger }\sigma _{-}\right) +\chi
\left( \hat{b}^{\dagger }\hat{a}^{2}+\hat{a}^{\dagger 2}\hat{b}\right) ,
\notag
\end{eqnarray}%
where $\sigma _{+}$($\sigma _{-}$) represents the raising and lowering
operators of the QD at the transition frequency $\omega _{0}$, and $\hat{a}%
^{\dagger }$($\hat{a}$) represents the creation and annihilation operators
of the fundamental mode with the resonance frequency $\omega _{a}$. Further,
$\hat{b}^{\dagger }$($\hat{b}$) represents the creation and annihilation
operators of the second harmonic mode with the resonance frequency $\omega
_{b}$. The driving field with the resonance frequency $\omega _{d}$ and
strength $E$ acts in the fundamental mode. The second harmonic mode is
driven by the pumping field with the resonance frequency $\omega _{p}$ and
strength $F$. Further, $g$ is defined as the coupling between the
fundamental mode and the QD, and $\chi $ is the parametric gain that denotes
the coefficient of second-order nonlinear interactions. In addition, $\delta
=\omega _{0}-\omega _{d}$ ($\Delta _{a}=\omega _{a}-\omega _{d}$) denotes
the detuning between the QD (fundamental mode) and the driving light, and $%
\Delta _{b}=\omega _{b}-\omega _{p}$ denotes the detuning between the second
harmonic mode and the pumping light.

Considering that the pumping light is very strong and the depletion by the
second harmonic mode is negligible, the equation of motion for the operator $%
\hat{b}$ can be approximately expressed as $id\hat{b}/dt=\left[ \hat{b},H%
\right] \simeq \left( \Delta _{b}-i\kappa _{b}/2\right) \hat{b}+F$, where $%
\kappa _{b}$ is the decay rate of the second harmonic mode. By allowing the
operator $\hat{b}$ to be time-independent, we obtain $\hat{b}\simeq F/\left(
i\kappa _{b}/2-\Delta _{b}\right) $. Substituting the expression of the
operator $\hat{b}$ in Eq. (\ref{H}), the effective Hamiltonian can be
reduced as follows,
\begin{equation}
H_{eff}=\delta \sigma _{+}\sigma _{-}+\Delta _{a}\hat{a}^{\dag }\hat{a}%
+g\left( \sigma _{+}\hat{a}+\sigma _{-}\hat{a}^{\dag }\right) +E\left( \hat{a%
}+\hat{a}^{\dagger }\right) +U\left( \hat{a}^{2}+\hat{a}^{\dagger 2}\right) ,
\label{Hamiltonian}
\end{equation}%
with the effective parametric gain $U=F\chi /\sqrt{\Delta _{b}^{2}+\kappa
_{b}^{2}/4}$. Here, we assumed real parametric gain via a well-chosen
relative phase of the pump field. Further, Eq. (\ref{Hamiltonian}) indicates
that the current model can be considered as a J-C model with a parametric
gain or bi-mode nonlinear cavity embedded with a single QD.

Unlike the J-C model, our scheme with the Hamiltonian in Eq. (\ref%
{Hamiltonian}) has an additional parametric gain. In addition, when $g=0$,
our scheme can be regarded as a bi-mode driven optical cavity, which has
been studied in Ref. \cite{14a} in which the PB arises only from the quantum
interference. As discussed below, we find the optimal condition of photon
antibunching in terms of effective parametric gain and coupling strength
based on the effective Hamiltonian $H_{eff}$, and analyze the difference
between the features of the PB effect and the J-C model and the bi-mode
driven optical cavity.
\begin{figure}[tbp]
\label{Fig1} \centering \includegraphics[width=1\columnwidth]{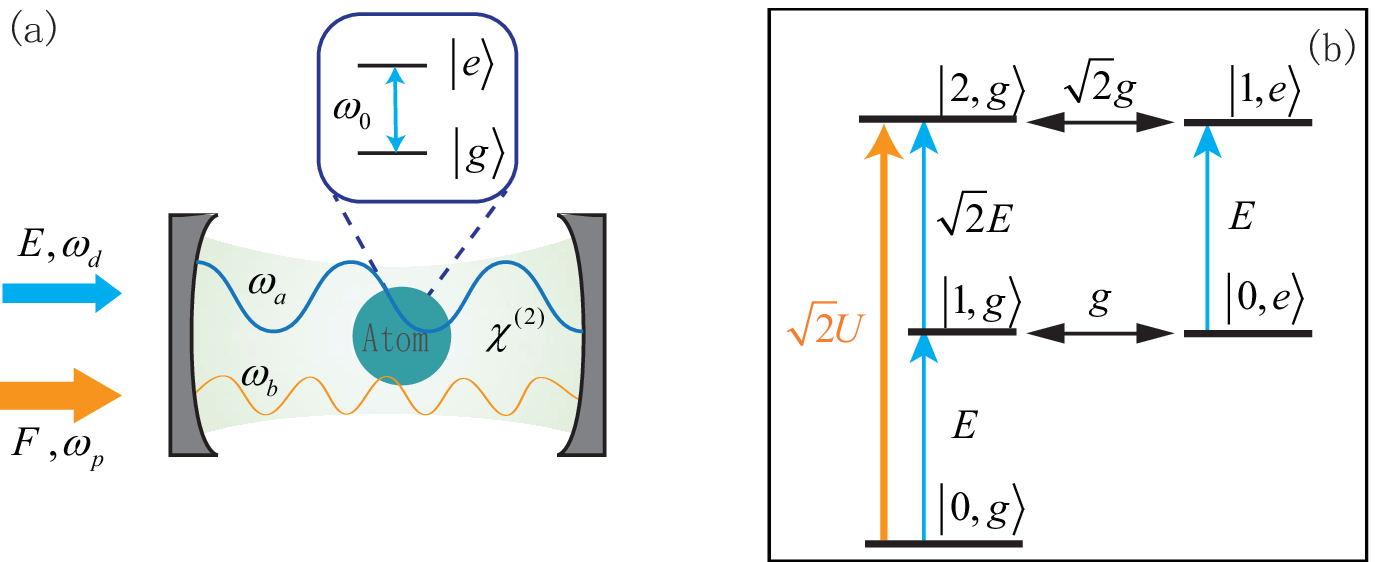}
\caption{(Color online) (a) Schematic showing a quantum dot embedded in a
nonlinear nanocavity made of $\protect\chi ^{\left( 2\right) }$ medium. (b)
Energy-level diagram showing the transition between zero-, one-, and two-
photon states and the transition paths leading to quantum interference for
obtaining antibunching photons.}
\end{figure}

\section{Optimal condition for photon antibunching}

It is known that $g^{\left( 2\right) }\left( 0\right) <1$ implies The
anti-bunched distribution of photons. In this section, we analytically
derive the optimal conditions for strong photon antibunching using the
wavefunction approach in Ref. \cite{our-work,13a,16,17}.

In a weak driving limit, under the assumption that higher photon excitation
states have a very low population, the wavefunction of the system is written
as,%
\begin{equation}
\left\vert \Psi \right\rangle =C_{0,g}\left\vert 0,g\right\rangle
+C_{0,e}\left\vert 0,e\right\rangle +C_{1,g}\left\vert 1,g\right\rangle
+C_{1,e}\left\vert 1,e\right\rangle +C_{2,g}\left\vert 2,g\right\rangle ,
\label{wavefunction}
\end{equation}%
where the coefficients $C_{n,g}$\ and $C_{n,e}$ denote the amplitudes of the
system in the states $\left\vert n,g\right\rangle $ and $\left\vert
n,e\right\rangle $, respectively. The state $\left\vert n,g\right\rangle $
(or $\left\vert n,e\right\rangle $) indicates that the system has $n$ $%
(n=0,1,2)$ photons of fundamental mode and is in the ground state $%
\left\vert g\right\rangle $ (or excited state $\left\vert e\right\rangle $)
of the atom. By using the wavefunction, we obtain $g^{\left( 2\right)
}\left( 0\right) \simeq 2\left\vert C_{2,g}\right\vert ^{2}/\left\vert
C_{1,g}\right\vert ^{4}$.

Now, substituting the wavefunction (\ref{wavefunction}) and the effective
non-Hermitian Hamiltonian in the Schrodinger equation, a set of equations
for coefficients can be given as follows:%
\begin{align}
i\dot{C}_{0,e}& =gC_{1,g}+\delta ^{^{\prime }}C_{0,e},  \label{1a} \\
i\dot{C}_{1,g}& =EC_{0,g}+gC_{0,e}+\Delta _{a}^{^{\prime }}C_{1,g},
\label{1b} \\
i\dot{C}_{1,e}& =\sqrt{2}gC_{2,g}+\left( \Delta _{a}^{^{\prime }}+\delta
^{^{\prime }}\right) C_{1,e}+EC_{0,e},  \label{1c} \\
i\dot{C}_{2,g}& =\sqrt{2}EC_{1,g}+\sqrt{2}gC_{1,e}+2\Delta _{a}^{^{\prime
}}C_{2,g}+\sqrt{2}UC_{0,g}.  \label{1d}
\end{align}%
Here, we defined $\delta ^{^{\prime }}=\delta -i\gamma /2$ and $\Delta
_{a}^{^{\prime }}=\Delta _{a}-i\kappa /2$ with $\gamma $ and $\kappa $ being
the decay rate of the QD and leakage rate of the fundamental mode of the
nanocavity, respectively. Using the above set of equations, we can obtain
the coefficients $C_{1,g}$ and $C_{2,g}$, when the system is in the steady
state ($\{\dot{C}_{n,e},\dot{C}_{n,g}\}=0$ ), as follows:
\begin{align}
C_{1,g}& =\frac{E\delta ^{^{\prime }}}{g^{2}-\delta ^{^{\prime }}\Delta
_{a}^{^{\prime }}},  \label{C1g} \\
C_{2,g}& =\frac{E^{2}\left[ g^{2}+\delta ^{^{\prime }}\left( \Delta
_{a}^{^{\prime }}+\delta ^{^{\prime }}\right) \right] -U\left( \delta
^{^{\prime }}\Delta _{a}^{^{\prime }}-g^{2}\right) \left( \Delta
_{a}^{^{\prime }}+\delta ^{^{\prime }}\right) }{\sqrt{2}\left[ \Delta
_{a}^{^{\prime }}\left( \Delta _{a}^{^{\prime }}+\delta ^{^{\prime }}\right)
-g^{2}\right] \left( \delta ^{^{\prime }}\Delta _{a}^{^{\prime
}}-g^{2}\right) }.
\end{align}%
Finally, the second order correlation function in the weak driving limit can
be presented:

\begin{eqnarray}
g^{\left( 2\right) }\left( 0\right) &\simeq &\frac{2\left\vert
C_{2,g}\right\vert ^{2}}{\left\vert C_{1,g}\right\vert ^{4}}  \label{g2} \\
&=&\left\vert g^{2}-\delta ^{^{\prime }}\Delta _{a}^{^{\prime }}\right\vert
^{2}\left\vert \frac{E^{2}g^{2}+\left( \Delta _{a}^{^{\prime }}+\delta
^{^{\prime }}\right) \left( E^{2}\delta ^{^{\prime }}-\delta ^{^{\prime
}}U\Delta _{a}^{^{\prime }}+Ug^{2}\right) }{E^{2}\delta ^{^{2\prime }}\left(
\Delta _{a}^{^{\prime }}\left( \Delta _{a}^{^{\prime }}+\delta ^{^{\prime
}}\right) -g^{2}\right) }\right\vert ^{2}.  \notag
\end{eqnarray}

Clearly, Eq. (\ref{g2}) indicates that, in the limit of strong coupling
strength, i.e., $g\gg (\gamma ,\kappa )$, when $\Delta _{a}\delta =g^{2}$ or
$g^{2}=\delta \left( \Delta _{a}+\delta \right) \left( U\Delta
_{a}-E^{2}\right) /\left( E^{2}+U\left( \Delta _{a}+\delta \right) \right) $%
, $g^{\left( 2\right) }\left( 0\right) $ takes local minimal values. We
further discuss the second order correlation function for both these cases,
where we first consider the case $\Delta \delta =g^{2}$. Under this
condition, the second order correlation takes a local minimum value as
follows,%
\begin{equation}
g^{\left( 2\right) }\left( 0\right) \approx \frac{\gamma ^{2}}{g^{2}}\left(
1+\frac{\gamma ^{2}U^{2}}{E^{4}}\right) .
\end{equation}%
It is noted that, if the effective parametric gain $U$ is significantly
smaller than the driving strength $E$, $g^{\left( 2\right) }\left( 0\right)
\ll 1$ corresponds to strong photon antibunching; therefore, $\Delta \delta
=g^{2}$ is a local optimal condition for photon antibunching. The physical
mechanism behind photon antibunching is that the strong interaction between
the photons blockades more photons entering the nanocavity at the same time,
which is the same as the case of photon blockade in the J-C model. The
optimal condition indicates that the strong antibunching induced by the
strong nonlinearity only occurs when the signs of atomic and cavity detuning
processes are the same.

On the contrary, if $C_{2,g}=0$, the second order correlation function
vanishes, i.e., $g^{\left( 2\right) }\left( 0\right) =0$. Based on Eq. (\ref%
{C2g}) we can arrive at the following condition to satisfy $C_{2,g}=0$:

\begin{equation}
g^{2}=\frac{U\Delta _{a}^{^{\prime }}-E^{2}}{E^{2}+U\left( \Delta
_{a}^{^{\prime }}+\delta ^{^{\prime }}\right) }\delta ^{^{\prime }}\left(
\Delta _{a}^{^{\prime }}+\delta ^{^{\prime }}\right) .  \label{Optimal1}
\end{equation}%
It is noted that Eq. (\ref{Optimal1}) is another local optimum condition for
strong photon antibunching. Unlike that in the strong coupling case, the
physical mechanism of the strong antibunching effect in this scenario
originates from quantum interference. From the energy level diagram (Fig.
1(b)), one observes that three different transition paths are available to
the two-photon state $\left\vert 2,g\right\rangle $: $\chi ^{\left( 2\right)
}$ nonlinearity interaction leading $\left\vert 0,g\right\rangle \overset{%
\sqrt{2}U}{\longleftrightarrow }\left\vert 2,g\right\rangle $; atomic-cavity
coupling inducing $\left\vert 1,e\right\rangle \longleftrightarrow
\left\vert 2,g\right\rangle $, and external field driving inducing $%
\left\vert 1,g\right\rangle \longleftrightarrow \left\vert 2,g\right\rangle $%
. If the system parameters satisfy the optimal condition, the total
transition results in a nearly perfect destructive interference and a nearly
vanishing population of the two-photon state $\left\vert 2,g\right\rangle $.
Moreover, we can find that, when $U\rightarrow 0$, our model reduces to J-C
model, and the optimal condition in Eq. (\ref{Optimal1}) reduces to $%
g^{2}=-\delta \left( \delta +\Delta _{a}\right) $, which is consistent with
the optimal condition for UCPB in the J-C model \cite{our-work}. On the
other hand, when $
g\rightarrow 0$, the system degenerates into a double-mode driven nonlinear
optical cavity, and the optimal condition (\ref{Optimal1}) becomes $%
E^{2}=U\Delta _{a}$, which is similar to the results of Ref. \cite{14a}.

\section{Numerical results}

We present the numerical results of the equal time second-order correlation
function $g^{(2)}(0)$ in this section to illustrate the photon statistics of
the fundamental mode. In numerical calculation, we evaluate $g^{\left(
2\right) }\left( 0\right) =Tr\left( \rho a^{\dagger }a^{\dagger }aa\right)
/Tr\left( \rho a^{\dagger }a\right) ^{2}$ by solving the master equation,
when the system is in the steady state; the master equation is defined as $%
\partial \rho /\partial t=-i\left[ H_{eff},\rho \right] +\kappa L\left[ \hat{%
a}\right] \rho /2+\gamma L\left[ \sigma _{-}\right] \rho /2$\ with $L\left[
\hat{o}\right] \rho =2\hat{o}\rho \hat{o}^{\dagger }-\hat{o}^{\dagger }\hat{o%
}\rho -\rho \hat{o}^{\dagger }\hat{o}$, denoting Lindblad equations.
Considering the system is in strong coupling, and taking $g=20\gamma $ as an
example, we numerically study and plot the dependence of the second order
correlation function $g^{(2)}(0)$ on cavity detuning $\Delta _{a}$ and QD
detuning $\delta $, as shown in Fig. 2(a). It can be seen from Fig. 2(a)
that, when the signs of cavity and QD detunings are the same (both $\delta $
and $\Delta _{a}$ are either blue or red), three troughs are present for
strong photon antibunching ($g^{(2)}(0)\ll 1$), which are represented by the
deep blue region. The expressions of two of these three troughs share the
same parametric condition with $\delta \Delta _{a}=g^{2}$ (highlighted by
red dashed line), which corresponds to the PB induced by strong nonlinear
interaction. One red dashed line is located in the first quadrant, and
another in the third quadrant, which results are consistent with those
discussed in Section III and with the case in a normal two-level atom
emitted cavity system\cite{our-work}. The remaining trough located in the
first quadrant, which is indicated by white dashed line with the numerical
optimal parametric relation, confirms the analytical results (Eq. (\ref%
{Optimal1})). It is noted that the PB induced by quantum interference under
this parametric condition-which is an interesting feature in our work-not
only originates from the strong nonlinear interaction but also is induced by
the quantum interference, when $\left\{ \delta ,\Delta _{a}\right\} >0$ with
$U\neq 0$. Apparently, one can see that the optimal condition in (\ref%
{Optimal1}) does not hold, when the cavity and atomic detunings (both in
red) are located in the third quadrant, implying that the photons are
bunched.

\begin{figure}[tbp]
\label{Fig2} \centering \includegraphics[width=1\columnwidth]{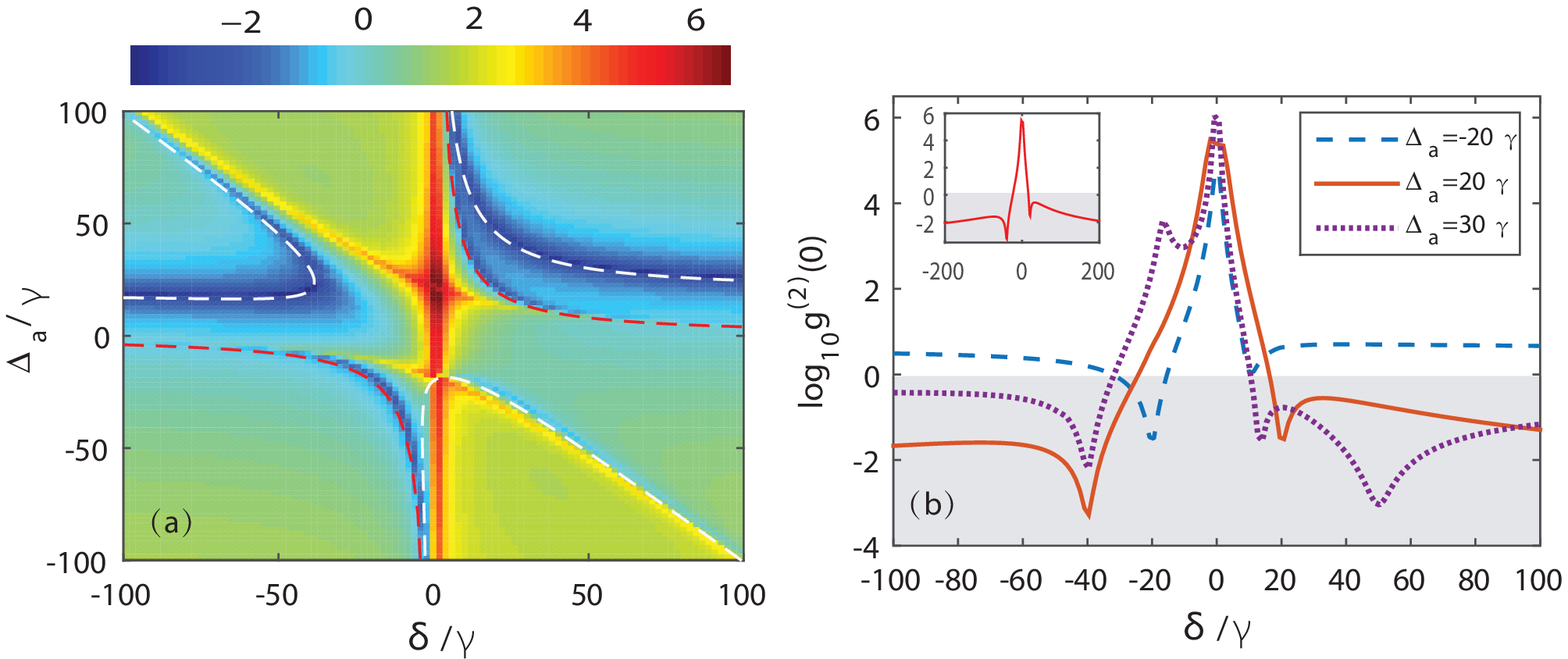}
\caption{(Color online) (a) Second-order correlation function log$%
_{10}g^{\left( 2\right) }\left( 0\right) $ as a function of atomic detuning $%
\protect\delta $ and cavity detuning $\Delta_a $. (b) Cross section of the
left panel along $\protect\delta $ with cavity detuning $\Delta_a =-20%
\protect\gamma ,20\protect\gamma ,30\protect\gamma $. Other parameters are
taken as $\protect\kappa =\protect\gamma $, $E=0.1\protect\gamma $, $U=0.0005%
\protect\gamma $, $g=20\protect\gamma $.}
\end{figure}

In addition, when the signs of cavity and QD detunings are opposite, $\delta
\Delta _{a}<0$. It is seen that one trough exists in $g^{(2)}(0)<1$ for
strong antibunching photons, in the second quadrant (QD detuning is
represented in red and cavity detuning in blue), indicated by the white
dashed line with the optimal parametric condition satisfying Eq. (\ref%
{Optimal1}), which corresponds to the PB originating from quantum
interference. However, for another condition of $\delta \Delta _{a}<0 $ (QD
detunings are represented in blue and cavity detuning in red), located in
the fourth quadrant, it is obvious that observing PB is difficult to observe
with respect to photon bunching. This scenario is completely different from
the cases in the first and the second quadrants, where the PB can be clearly
observed.

Then, we plot $g^{(2)}(0)$ as a function of atomic detuning $\delta $ in
Fig. 2(b). It is seen that, when cavity detuning is in the red region,
taking $\Delta _{a}=-20\gamma $ as an example, only one trough is located at
$\delta =-20\gamma $ with red QD detuning; the corresponding optimal value
of second correlation function is $g^{(2)}(0)\approx 0.022$. In this case,
we found that both the cavity and QD detunings are in red with the optimal
parametric condition satisfying $\delta \Delta _{a}=g^{2}$, which
corresponds to the PB arising from the strong nonlinear interaction. This
result indicates that only one feasible method of strong nonlinear
interaction is available to observe PB, when $\Delta _{a}<0$.

Furthermore, when the cavity detuning is red, we take $\Delta _{a}=20\gamma
,30\gamma $ as examples. When $\Delta _{a}=20\gamma $, i.e., $\Delta
_{a}=E^{2}/U$, we find that there are only two troughs in $g^{(2)}(0)\ll 1$
for stronger antibunching photons. A trough at $\delta \simeq 20\gamma $,
whose parametric condition satisfies $\delta \Delta _{a}=g^{2}$, is
consistent with the case when $\Delta _{a}=-20\gamma $, which corresponds to
the PB originating from the strong nonlinear interaction, with both cavity
and QD detunings being blue. Another trough located at $\delta =-40\gamma
=-2E^{2}/U$, whose optimal parametric condition for strong antibunching is
given by Eq. (\ref{Optimal1}), corresponds to the PB caused by quantum
interference. A comparison of the values of $g^{(2)}(0)$ in these two
troughs indicates that $g^{(2)}(0)$ in $\delta =-40\gamma $ is smaller than $%
g^{(2)}(0)$ in $\delta =20\gamma $, implying stronger photon antibunching,
when PB is caused by quantum interference. Now, we discuss another case in
which$\Delta _{a}=30\gamma >E^{2}/U$; here, one can see that, in addition to
$g^{(2)}(0)\ll 1$ for strong antibunching photons at $\delta =-40\gamma $, $%
\delta =50\gamma $ exists, from which the PB originating from quantum
interference is observed as well; both the parameters given here satisfy Eq.
(\ref{Optimal1}). Unlike the case with $\Delta _{a}=E^{2}/U$, two troughs
are present from which the PB effect is observed; the PB in both these cases
was induced by quantum interference. Apparently another trough is located at
$\delta \approx -13.3\gamma $ for the PB effect induced by strong nonlinear
interaction. Apart from the above discussion, we can see that $g^{(2)}(0)$
tends to stabilize at $\left\vert \delta \right\vert >2E^{2}/U$, as
illustrated in Fig. 2(b) with $\Delta _{a}=20\gamma $.

\begin{figure*}[tbp]
\label{Fig3} \centering \includegraphics[width= 1\columnwidth]{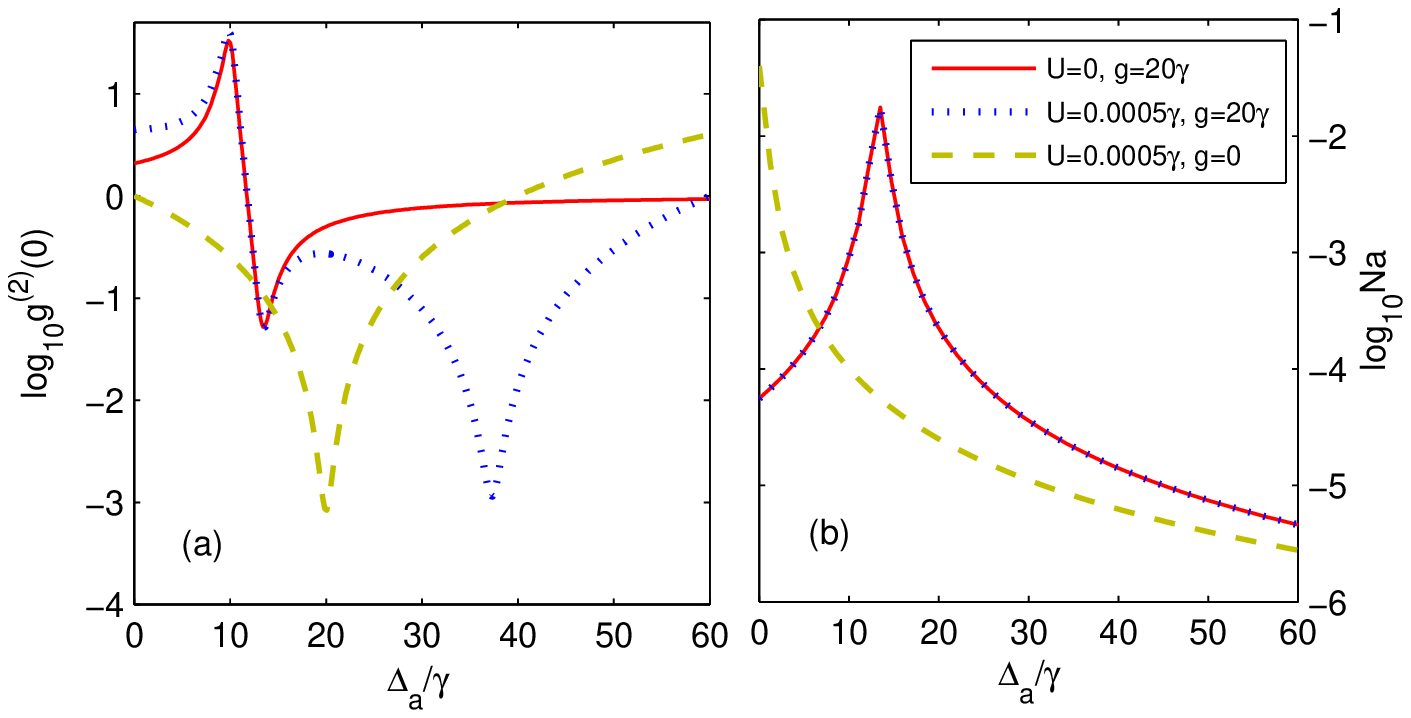}
\caption{(Color online) Second-order correlation function log$_{10}g^{(2)}(0)
$ and average photon number log$_{10}N_{a}$ as functions of cavity detuning $%
\Delta_a$ with red solid line on $U=0$ and $g=20\protect\gamma $, purple
dotted line on $U=0.0005\protect\gamma $ and $g=20\protect\gamma $, yellow
dashed line on $U=0.0005\protect\gamma $, and $g=0$ in (a) and (b)
respectively. Other parameters are taken as $\protect\kappa =\protect\gamma $%
, $E=0.1\protect\gamma $, $\protect\delta=30\protect\gamma $.}
\end{figure*}

The above discussion clearly indicates that the model proposed in this paper
under different parameters can be reduced to a different well-known model.
For example, when $U=0$, our model is reduced to J-C model, and when $g=0$,
it becomes a bi-mode driven optical cavity. To further understand the photon
statistical properties of these three models under the same parameters, we
plot $g^{(2)}(0)$ with cavity detuning $\Delta _{a}$ in Fig. 3(a) with QD
detuning $\delta =30\gamma $. The red line represents the case where $U=0$
and $g=20\gamma $, corresponding to the J-C model. One can see that a trough
exists, where the minimum value of $g^{(2)}(0)$ is significantly less than $1
$, implying a strong photon antibunching. The trough is located at $\Delta
_{a}=13.3\gamma $. Obviously, the parameter satisfying the optimal relation $%
\delta \Delta _{a}=g^{2}$ and photon antibunching comes from the strong
nonlinear interaction. The yellow dashed line indicates the condition of a
bi-mode driven optical cavity with the parameters $g=0$ and $U=0.0005\gamma $%
. A highly deep trough is found at $\Delta _{a}=20\gamma $, where $%
g^{(2)}(0)\approx 0.001$, which is greatly lower than that of the J-C model.
Now, the parameters satisfy the optimal parametric relation $\Delta
_{a}U=E^{2}$. Clearly, the strong photon antibunching at the optimal
parameter point is caused by quantum interference.

In addition, we study the case with existing parametric gain and a QD in the
cavity by setting the parameters as $U=0.0005\gamma $ and $g=20\gamma $.
Now, the system can be regarded as a composite system of the J-C model and
the bi-mode driven optical cavity. The result is represented by the blue
dotted line in Fig. 3(a). It is noted that two troughs are present at $%
g^{(2)}(0)\ll 1$ in the red QD detuning region. The first is located at $%
\Delta _{a}=13.3\gamma $, satisfying the optimal parameter relation $\delta
\Delta _{a}=g^{2}$. Thus, the physical mechanism of PB arises from the
strong nonlinear interaction, similar to the J-C model. This result implies
that the parametric gain does not affect the nonlinear interaction between
the QD and the fundamental mode of cavity. The second trough is located at $%
\Delta _{a}=37.3\gamma $, which satisfies the optimal parameter relation (%
\ref{Optimal1}), indicating that quantum interference causes PB. However,
the optimal cavity detuning in the composite system differs from that in the
bi-mode driven optical cavity. The principle behind this phenomenon is that,
compared with the bi-mode driven optical cavity, three transition paths,
rather than two, are involved, resulting in a destructive quantum
interference in the composite system, as shown in Fig. 1(b). The above
discussion illustrates that the fundamental mode in the proposed composite
system has a more comprehensive quantum statistics behavior than that in the
J-C mode and the bi-mode driven optical cavity.

Finally, we discuss the influence of quantum interference on the average
photon number. The average photon number of the fundamental mode of the
cavity is $n_{a}=\left\langle a^{\dagger }a\right\rangle \simeq \left\vert
C_{1g}\right\vert ^{2}$. Comparing Eq. (\ref{C1g}) and Eq. (14) in Ref.\cite%
{our-work}, one can see that the average photon number in the proposed model
is the same as that in the J-C model, which is confirmed by the numerical
results in Fig. 3(b) as well. The blue dotted line in Fig. 3 indicates that,
for the proposed model, although the quantum interference between the two
transition pathways greatly influences the photon statistics and results in
a strong photon antibunching around $\Delta _{a}=37.3\gamma $, the average
photon number remains almost unaffected. Therefore, in the weak driving
limit, the average photon number is almost determined by the population of
the single-photon state $\left\vert 1,g\right\rangle $, and the second order
correlation function is approximately proportional to the two-photon state $%
\left\vert 2,g\right\rangle $. The quantum interference affects only the
two-photon state $\left\vert 2,g\right\rangle $ and therefore the quantum
statistics, rather than the average photon number of the fundamental mode.
The second order correlation function and the average photon number of the
bi-mode driven cavity (yellow dashed lines in Fig. 3) further corroborate
this conclusion.

\section{Conclusion}

In this study, we proposed a physical model to realize
conventional/unconventional photon blockade in a parametric driven
nanocavity embedded with a quantum dot. In the weak driving limit, we used
the wave function method to obtain the second-order correlation function
expression of the fundamental mode of the nanocavity, and found the optimal
parametric conditions to realize strong photon blockade. By analyzing the
second order correlation function, we found that conventional and
unconventional photon blockades exist simultaneously in the red cavity
detuning region. By comparison with the J-C model and the bi-mode driven
nanocavity model, we found that the photon blockades originating from strong
nonlinearity and quantum interference exist simultaneously in the proposed
model. In addition, we found that quantum interference does not influence
the average number of photons, whereas it greatly influences the statistical
distribution of photons.

\section*{Funding Information}

National Natural Science Foundation of China (NSFC) (11664014,11504145,
11964014, and 11364021).

\end{document}